# On Explaining Recommendations with Large Language Models: A Review

**Alan Said**

*University of Gothenburg, Gothenburg, Sweden*

Correspondence*:
Corresponding Author
alan@gu.se

## ABSTRACT

The rise of Large Language Models (LLMs), such as LLaMA and ChatGPT, has opened new opportunities for enhancing recommender systems through improved explainability. This paper provides a systematic literature review focused on leveraging LLMs to generate explanations for recommendations — a critical aspect for fostering transparency and user trust. We conducted a comprehensive search within the ACM Guide to Computing Literature, covering publications from the launch of ChatGPT (November 2022) to the present (November 2024). Our search yielded 232 articles, but after applying inclusion criteria, only six were identified as directly addressing the use of LLMs in explaining recommendations. This scarcity highlights that, despite the rise of LLMs, their application in explainable recommender systems is still in an early stage. We analyze these select studies to understand current methodologies, identify challenges, and suggest directions for future research. Our findings underscore the potential of LLMs improving explanations of recommender systems and encourage the development of more transparent and user-centric recommendation explanation solutions.

## 1 INTRODUCTION

Recommender systems have become integral to various digital platforms, assisting users in navigating vast amounts of information by suggesting products, services, or content tailored to their preferences. From e-commerce sites recommending products to streaming services suggesting movies, these systems aim to enhance user engagement and satisfaction. However, as recommender systems grow in complexity, users often remain unaware of the underlying mechanisms that generate these personalized suggestions. Eslami et al. (2019) and Ge et al. (2024) identified that this opacity can lead to mistrust, reduced satisfaction, reduced engagement, and reluctance to adopt new recommendations. Engagement in this context refers to the users' willingness to continue interacting with, or using, the system. Explanations within recommender systems serve as a bridge between complex algorithms and user understanding. By providing insights into why a particular item is recommended, explanations empower users to make informed decisions, enhancing their overall experience. They help users assess the relevance of recommendations, increase transparency, and build trust in the system (Lu et al., 2023). For instance, an explanation that highlights how a recommended book aligns with a user's past reading habits can make the suggestion more persuasive and acceptable.

Moreover, explanations can alleviate concerns about privacy and data usage by clarifying how user information is used when generating recommendations (Abdollahi and Nasraoui, 2018). They contribute to a sense of control, allowing users to adjust their preferences and interact more effectively with the system.

In scenarios where recommendations might seem unexpected or irrelevant, explanations can mitigate confusion and prevent user disengagement (Tintarev and Masthoff, 2022).

There are two primary categories of explainable AI methods commonly used in recommender systems: local and global explanations. Local explanations focus on providing insights into a specific recommendation, explaining why a particular item was suggested to a particular user. In contrast, global explanations aim to provide a holistic understanding of the overall behavior of the recommender model, offering insights into how the model functions generally across all users. Additionally, methods for explainability can be categorized as model-specific or model-agnostic. Model-specific methods are designed for specific types of algorithms, while model-agnostic methods can be applied to any machine learning model. Common frameworks like LIME (Local Interpretable Model-agnostic Explanations) by Ribeiro et al. (2016) and SHAP (SHapley Additive exPlanations) by Lundberg and Lee (2017) are widely used to generate such explanations. These frameworks provide mechanisms to explain individual predictions (local explanations) or give a broader view of the model's decision-making process (global explanations). Explanations based on these methods provide a sense of technical transparency, as they generate explanations based on the features of a specific item or algorithnic model.

While explanations generated by LLMs such as, e.g., OpenAI's ChatGPT (Brown et al., 2020) or Meta's LlaMA (Touvron et al., 2023) offer new opportunities for generating rich, natural language descriptions of why an item is recommended, they do not necessarily conform to traditional local or global explanation frameworks. Unlike LIME and SHAP, LLMs focus on producing contextually relevant, human-readable interpretations of recommendations. These interpretations tend to *justify* why a recommendation might make sense from a user's perspective rather than *explain* by analyzing the algorithm's internal mechanics. In this sense, LLMs thus offer accessible justifications rather than precise, analytic explanations (as discussed by Silva et al. (2024)).

This literature review explores the intersection of recommender systems, explanations, and large language models. By examining recent studies that leverage LLMs to generate explanations for recommendations, we aim to understand the current level of integrating LLM-based explanations (or justifications), identify challenges, and highlight opportunities for future research.

## 2 METHODOLOGY

For data collection, we adhered to the systematic literature review procedure outlined in the guidelines by Kitchenham et al. (2007). To develop an effective search strategy, we conducted a scoping review of relevant published literature. The scoping review procedure is outlined in Fig. 1. In line with findings by Bauer et al. (2024), this scoping review revealed that the keywords *recommendation systems* and *recommender systems* are used interchangeably, with the latter being more prevalent in the research community centered around the ACM Conference on Recommender Systems (RecSys)[1], while both terms are widely used in other research outlets.

### 2.1 Literature Search

With our aim to cover research revolving around explainable recommender systems, we identified that searches using the keywords *explainability* and *explanations* significantly overlap. To ensure the inclusion of comprehensive and substantial studies, we limited our search to works labeled as *Research Article* or *Short Paper*, thereby excluding abstracts, reports, and similar publications.

[1] https://recsys.acm.org/

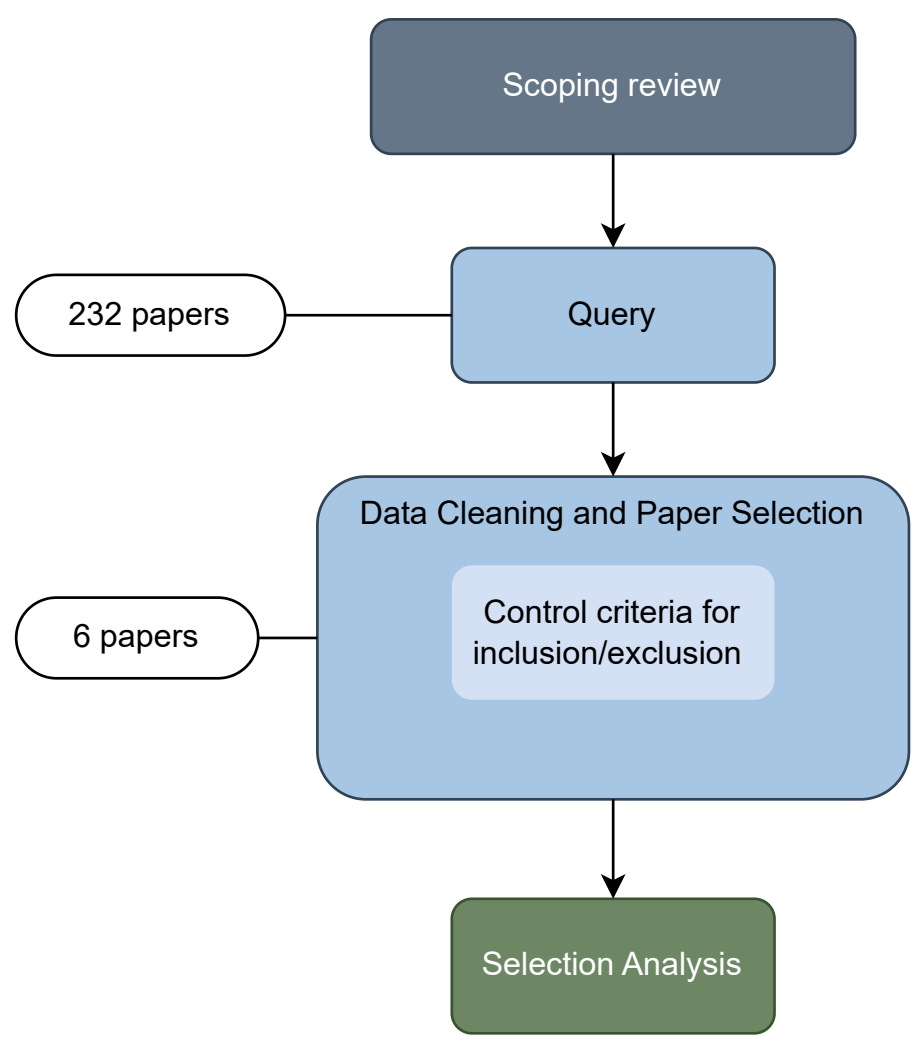

**Figure 1.** Paper search and selection procedure for the surveyed papers.

Our search strategy to identify eligible papers consisted of several consecutive stages. The ACM Digital Library[2] not only contains papers published by ACM but also by other publishers, allowing us to search for papers in the main established conferences and journals where research on recommender system evaluation is published. Besides the main conference on recommender systems—ACM RecSys—this includes conferences such as SIGIR, CIKM, UMAP, and KDD. Relevant journals include, for instance, *TORS*, *TOIS*, *TIST*, and *TIIS*.

Accordingly, we sampled papers from the ACM Digital Library (The ACM Guide to Computing Literature), which is described as "*the most comprehensive bibliographic database in existence today focused exclusively on the field of computing.*"[3]

We considered papers within an encapsulated time frame from November 1, 2022 (the launch of ChatGPT), to the present (November 1, 2024 at the time of writing), assuming that the employed databases and search engines have already completed indexing papers from conferences and journals.

As our literature review focuses on research regarding the explanation of recommender systems using large language models, we searched for papers indexed with the terms *recommendation system** or *recommender system** (covering both commonly used terms), and *large language model**, *LLM**, *ChatGPT*, or *Chat-GPT* (to encompass related LLM terms), as well as the term *expla** (to cover *explainability* and *explanation(s)*).

Altogether, this resulted in the following query:

[2] https://dl.acm.org

[3] https://libraries.acm.org/digital-library/acm-guide-to-computing-literature

```
        "query":
            {
                Fulltext:(
                    "recommender system*"
                    OR
                    "recommendation system*"
                )
                AND
                Fulltext:(
                    "large language model*"
                    OR
                    "llm*"
                    OR
                    "chatgpt"
                    OR
                    "chat-gpt"
                )
                AND
                Fulltext:(expla*)
            }
            "filter": {
                Article Type: Research Article OR Short Paper,
                E-Publication Date: (11/01/2022 TO 11/01/2024)
            }
```

This query[4] returned a total of 232 hits as of November 5, 2024.

## 2.2 Literature Selection

A paper was included if it fulfilled *all* of the following criteria (pre-established inclusion criteria):

(i) The paper deals with the *explanation of recommendations*.
(ii) The paper uses a *large language model for generating explanations*.
(iii) The paper is a *Research Article* or a *Short Paper*.
(iv) The paper was *published within the time range from November 1, 2022, to November 1, 2024, inclusive*.

A paper was excluded if it met *any* of the following criteria (pre-established exclusion criteria):

(a) The paper is not a research article or short paper.
(b) The paper is an abstract, demo paper, tutorial paper, or report.
(c) The paper is a literature review.

[4] https://dl.acm.org/action/doSearch?fillQuickSearch=false&target=advanced&expand=all&AfterMonth=11&AfterYear=2022&BeforeMonth=10&BeforeYear=2024&AllField=Fulltext%3A%28%22recommender+system*%22+OR+%22recommendation+system*%22%29+AND+Fulltext%3A%28%22large+language+model*%22+OR+%22llm*%22+OR+%22chatgpt%22+OR+%22chat-gpt%22%29+AND+Fulltext%3A%28expla*%29&startPage=&ContentItemType=short-paper&ContentItemType=research-article NB: Due to how ACM indexes papers from other publishers, the number of retrieved papers may differ on different dates.

(d) The paper is not concerned with recommender systems.
(e) The paper does not contribute to explanations of recommendations.

This data cleaning and selection procedure led to the exclusion of 226 papers. The remaining six papers constitute our final sample resulting from the query. Table 1 provides an overview of the papers in the sample, including venue where it was published, paper type, and reasons for inclusion.

**Table 1.** Overview of the included papers, ordered by year and venue.

| Paper | Venue | Type | Reason for inclusion |
|---|---|---|---|
| Park et al. (2023) | EICS | Late-breaking Results | Conversational explanations |
| Guo et al. (2023) | SIGIR | Resource Paper | Evaluations of explanations |
| Silva et al. (2024) | IUI | Research Paper | LLM-based explanations |
| Petruzzelli et al. (2024) | RecSys | Research Paper | LLM-based cross-domain recommendation and explanations |
| Hendrawan et al. (2024) | ExUM @ UMAP | Workshop paper | LLM-based explanations |
| Lubos et al. (2024) | ExUM @ UMAP | Workshop paper | LLM-based explanations |

### 2.3 Literature Analysis

The systematic search identified only six papers focusing on recommendation explanations using LLMs. We analyzes these papers focusing on use-cases, datasets, methodologies, and research questions investigated in the context of these works in order to get an understanding of the current state of the art of leveraging LLMs for recommendation explanations. The majority of the excluded papers mentioned recommendation and explanation in the context of LLMs, albeit without applying LLMs for the purpose of generating explanations. The bulk of the retrieved 232 papers were published in 2024 (186), notably, only two out of the retrieved papers were published in 2022 pointing to the recent rapid growth of the LLM-related literature.

## 3 RESULTS

After applying the specific inclusion and exclusion criteria for this literature review (cf. Section 2), only six papers were selected for detailed analysis. Given that less than two years have passed since the launch of ChatGPT, the publication and peer review of these six papers at conferences and workshops signal a rapidly emerging field. The selected papers were published at EICS, SIGIR (both 2023), and IUI, ExUM@UMAP, and RecSys (all 2024) reflecting the increasing focus the potential of LLMs for generating explanations in recommender systems.

### 3.1 Papers

Below, we provide a detailed overview of each of the six included papers.

#### 3.1.1 A User Preference and Intent Extraction Framework for Explainable Conversational Recommender Systems (Park et al., 2023)

The paper focuses on improving conversational recommender systems (CRS) by addressing two primary challenges, i.e., preference extraction from users during conversations and transparency in recommendations. Traditional CRSs have limitations in extracting accurate and context-aware user preferences and often function as black-box models, leading to a lack of transparency in their recommendations. To overcome these issues, the authors propose a novel framework that combines entity-based user preference extraction with context-aware intent extraction, generating recommendations that are both interpretable and aligned with user preferences. The framework uses item feature entity detection and sentiment analysis to extract user preferences and assigns ratings to each feature mentioned during the conversation. It then uses graph-based representations of user utterances to capture contextual user intent, enhancing the accuracy of recommendations. The final recommendation phase involves candidate selection and ranking without the use of black-box models, ensuring the system is explainable. For generating explanations, the paper uses GPT-2 (Radford et al., 2019), which provides users with natural language explanations based on item features and estimated user preferences for the recommendations.

The framework was evaluated using two datasets, INSPIRED introduced by Hayati et al. (2020) and ReDial by Li et al. (2018), demonstrating improvements in preference extraction and recommendation accuracy compared to existing methods. The paper concludes by highlighting the potential of this approach to make CRSs more transparent and user-friendly, while also outlining future research directions, such as conducting user studies to evaluate satisfaction with the explanations.

### 3.1.2 Towards Explainable Conversational Recommender Systems (Guo et al., 2023)

This paper focuses on improving explainable CRSs by addressing the challenges associated with generating multiple contextualized explanations in real-time. It proposes a novel set of ten evaluation perspectives to measure the quality of explanations in CRS, such as effectiveness, persuasiveness, and trust. To address the lack of high-quality explanations in existing datasets, the authors develop E-ReDial, a new conversational recommendation dataset with over 2,000 manually rewritten high-quality explanations for movies.

The paper highlights two baseline approaches for generating explanations in CRS, i.e., training-based models, such as T5 (Raffel et al., 2020), and prompt-based models, such as GPT-3 (Brown et al., 2020), evaluated on their ability to generate explanations using the newly constructed dataset. The results show that models trained on E-ReDial outperform existing methods in producing explanations that are clearer, more persuasive, and contextually appropriate. Additionally, integrating external knowledge, such as movie reviews and descriptions, further improves performance, especially in terms of explanation quality.

The results demonstrate the effectiveness of E-ReDial in improving CRS explainability, while also pointing out future directions for improving explainable CRS, including the need of developing more realistic datasets and advancing automatic explanation generation techniques.

### 3.1.3 Leveraging ChatGPT for Automated Human-centered Explanations in Recommender Systems (Silva et al., 2024)

This paper investigates the use of ChatGPT to generate personalized, human-centered explanations in recommender systems. The paper aims to address the need for transparency and interpretability in RS, which the paper states are crucial for building user trust. Current systems often lack scalable and meaningful ways to explain their recommendations. To fill this gap, the paper proposes leveraging ChatGPT to generate personalized explanations for movie recommendations, with a focus on user experience.

The paper presents a user study evaluating ChatGPT's ability to generate explanations of recommendations, involving 94 participants. Each participant provided movie preferences, and ChatGPT was tasked with generating both personalized recommendations and disrecommendations (items to avoid). The study assessed whether users preferred ChatGPT-generated recommendations and explanations over random (but popular) recommendations.

The paper's key findings state that users preferred ChatGPT-generated recommendations over random selections of popular movies, that personalized explanations based on user preferences were not perceived as significantly more effective or persuasive than generic ones unless the recommendations were random, and that disrecommendations also benefited from ChatGPT's explanations, although generic explanations often outperformed user-based ones in these cases.

The paper concludes that while ChatGPT can effectively generate natural language explanations (referred to as justifications), more work is needed to fully understand how users perceive personalization and what makes explanations truly persuasive. The paper also explores how explanation goals, such as effectiveness and persuasiveness, relate to one another through a path model analysis, revealing that persuasiveness plays a key role in how users judge explanation quality.

#### 3.1.4 Explanations in Open User Models for Personalized Information Exploration (Hendrawan et al., 2024)

This paper explores how open user models can enhance transparency and user control in personalized information exploration systems. The focus is on providing users with control over the recommendation process by allowing them to adjust their profile through adding or removing topics and receiving explanations for these choices. The study is conducted within the Grapevine system by Rahdari et al. (2020), an exploratory search platform designed to help students find faculty advisors for research projects. The paper extends the system by incorporating LLMs to generate explanations, introducing two types of explanations: individual keyphrase explanations and relationship explanations between keyphrases. These explanations are intended to assist users in understanding the significance of the topics in their profile and how different topics are interconnected.

The study features an observational experiment with 23 participants, who used the Grapevine system to select research topics and faculty members. The results show that users who accessed explanations, especially those that explored relationships between topics, exhibited higher engagement with the system and made more refined adjustments to their user profile. However, it was noted that while individual explanations improved users' confidence, relationship explanations were used less frequently and led to mixed satisfaction levels.

The paper highlights the potential benefits of using LLMs to generate contextual explanations in personalized systems but also points out challenges, such as the need for clearer explanations and more intuitive interfaces to improve user satisfaction.

#### 3.1.5 LLM-generated Explanations for Recommender Systems (Lubos et al., 2024)

The paper explores the use of LLMs, such as LlaMA 2, to generate personalized explanations for three types of recommender systems: feature-based, item-based collaborative filtering, and knowledge-based recommendations. It examines the potential of LLMs to improve the quality of explanations through natural language generation and evaluates how users perceive these explanations compared to traditional methods.

To gather insights, the authors conducted an online user study with 97 participants, who interacted with LLM-generated explanations across different recommendation types. The study aimed to explore user preferences, assess the quality of LLM-generated explanations, and identify characteristics that users found appealing. Results indicated a clear preference for explanations generated by LLMs over baseline approaches, with participants appreciating the detail, creativity, and informative nature of the explanations. Users favored LLM-generated explanations for their ability to provide contextual information and background knowledge, which enhanced trust to and satisfaction with the recommendations.

However, the study also identified challenges, particularly in maintaining clarity for more complex explanations, such as those based on knowledge-based recommendations. The paper highlights the need for further research on LLM use and integratiion of external information into LLM-generated explanations to handle more specialized domains effectively.

### 3.1.6 Instructing and Prompting Large Language Models for Explainable Cross-domain Recommendations (Petruzzelli et al., 2024)

The paper explores using LLMs to generate explainable recommendations in cross-domain recommendation tasks, focusing on both improving recommendation accuracy and producing natural language explanations. Through instruction-tuning and personalized prompting, LLMs like GPT, LLaMa, and Mistral (Jiang et al., 2023) generate contextual explanations alongside recommended items, offering personalized justifications based on users' preferences in a source domain. This strategy allows the LLM to deliver coherent and relevant explanations, addressing transparency issues that commonly arise in cross-domain settings.

Experimental results show that LLM-produced explanations are not only readable but also contextually rich, revealing how past preferences influence new recommendations. The study highlights the potential of in-context learning to further enhance the relevance and quality of explanations, demonstrating that LLMs can provide intuitive, user-friendly justifications that build user trust and understanding in cross-domain recommendations.

## 4 DISCUSSION

The reviewed papers collectively study how LLMs can be used to enhance the explainability of recommender systems, revealing several insights into LLM-generated explanations' impact on user experience. Across the studies, there is a trend toward shifting from traditional, rigid explanation methods—such as feature-based or collaborative filtering explanations—to more flexible and personalized natural language explanations offered by LLMs. Silva et al. (2024), for example, showed that ChatGPT improved user engagement through human-centered, contextually relevant explanations. Similarly, Lubos et al. (2024) found that users generally preferred LLM-generated explanations for their creativity and depth, which enhanced trust and satisfaction in recommendations.

A common challenge identified with LLM-generated explanations is balancing detail with clarity. While detailed explanations can enrich user experience, as observed in Hendrawan et al. (2024), they can sometimes overwhelm users, particularly in specialized or knowledge-intensive domains. This underscores the importance of avoiding unnecessary complexity and maintaining explanation clarity.

The papers also explore the unique capabilities of LLMs in generating cross-domain recommendations. For example, Petruzzelli et al. (2024) demonstrated how personalized prompting can be used to generate relevant explanations that connect user preferences from a source domain to recommendations in a target domain. This cross-domain adaptability highlights the potential of in-context learning to improve the relevance and quality of explanations across different recommendation settings, emphasizing the versatility of LLMs in handling diverse recommendation challenges.

An important distinction made in several studies is between generating *explanations* and *justifications*. LLMs, as noted by Park et al. (2023) and Guo et al. (2023), bypass traditional model-explainability frameworks, and instead on produce accessible justifications that improve perceived transparency. This approach emphasizes *perceived* rather than *technical* transparency, which aligns with the goal of enhancing user experience but may limit the depth of introspection provided into the model's workings.

Lastly, the studies identify a need for evaluation methods and datasets tailored to LLM-generated explanations (Guo et al., 2023). The continued integration of LLMs into recommender systems will likely benefit from systematic assessment methods of explanation quality and effectiveness across varied domains and user needs.

## 5 CONCLUSIONS

This review demonstrates the potential of LLMs to deliver user-friendly explanations in recommender systems, offering a level of personalization and contextual richness that traditional approaches lack. While these models excel at generating natural language justifications, they shift focus from technical transparency to enhancing perceived transparency for users. As a result, LLMs contribute to a more engaging and accessible explainability, though they do not provide the algorithmic introspection that formal explanation frameworks offer. Leveraging their linguistic capabilities, LLMs open new possibilities for personalizing and tailoring explanations, not only in terms of which insights to include but also in how these insights are conveyed to enhance user satisfaction. Furthermore, LLMs allow for incorporating external information—such as sustainability factors—broadening the scope of explanations to better align with user values and preferences.

Looking forward, one possible direction is to refine LLM-generated explanations to ensure they remain clear and concise, especially in complex recommendation contexts. Developing approaches that filter and summarize information without losing relevance is central for avoiding user overload. Another promising research direction is combining LLMs with traditional explanation methods methods to provide hybrid explanations that blend intuitive, human-readable justifications with local or global insights. This dual approach could support both casual users seeking quick understanding and more technically inclined users wanting a deeper explanation of the recommendation process.